\documentclass{aa}
\usepackage{graphics} 
\usepackage{epsf} 
\usepackage{epsfig}
\usepackage{psfig}
\usepackage{amssymb}
\usepackage{txfonts}
\usepackage{rotating}

\begin{document}


\title{Late-type Giant Variables in NGC 6522, LMC and SMC \\How do they differ?}

\titlerunning{Long-Period Variables in NGC 6522, LMC and SMC}

\author{M. Schultheis \inst{1,2}
\and I.S. Glass \inst{3} 
\and M.-R. Cioni\inst{4}}



\institute{CNRS UMR 6091, Observatoire de Besan\c{c}on, BP1615, F-25010 Besan\c{c}on Cedex France     
\and   CNRS UMR 7095, Institut d'Astrophysique de Paris, 98bis, bld. Arago, 75014 Paris, France  
\and South African Astronomical Observatory, P,O. Box 9, Observatory 7935, South Africa
\and  European Southern Observatory, Karl-Schwarschild-Str. 2, 85748 Garching bei M\"unchen, Germany 
}  

\offprints{mathias@obs-besancon.fr} 


\date{Received ....  / Accepted .. .........}  



\abstract{
Complete samples of 2MASS stars in three fields of differing metallicities
(and possibly differing age distributions) have been extracted and
cross-correlated with MACHO and ISO data to determine their variability and
mass-loss properties.\\ In each $M_K$ histogram a fall-off in numbers is
seen above the tip of the RGB. The luminosity of the tip increases with
metallicity as does the average $(J-K)_0$ at a given $M_{K_0}$. These
quantities have been compared with the data for galactic globular clusters
obtained by Ferraro et al. (\cite{Ferraro2000}). In the $J-H$, $H-K$
colour-colour diagrams, the increasing proportion of stars with high $H-K$
values is conspicuous at lower metallicities. This is well-known to result
from the increasing proportion of carbon stars.

All three fields contain similar types of variables, mainly short-period
(usually tens of days), Mira like (long-period, large amplitude) and
double-period (having both long and short periods). The proportion of stars
that vary decreases at lower metallicities and the minimum period associated
with a given amplitude gets longer. Various trends are seen in the $K$,
log$P$ diagrams of each field. The Magellanic Cloud fields largely resemble
each other but the Bulge field is noticeably different. The stars that
follow the `A' $K$, log $P$ relation in the Bulge hardly extend beyond
$M_{K,0}^{\rm Tip}$ and the other groups also appear truncated in $M_K$
relative to their Magellanic Cloud counterparts. In the Clouds there are
many stars with periods around 200-300d that follow the `C' or Mira relation
but have relatively small amplitudes.

The mid-IR sample detected by ISOCAM during the Magellanic Cloud Mini-Survey
(MCMS) appears to be reliable and complete for sources with $M_K$ more
luminous than -- 7 mag, i.e., for those close to the top of the AGB.
The various colour-colour and colour-magnitude diagrams reflect the
increasing dominance of carbon stars at low metallicity. Mira magnitude vs
log $P$ relations exist at least up to 7$\mu$m.  Mass-loss from
longer-period and double-period SRVs occurs at similar rates in each field,
in spite of the metallicity differences.
}

\maketitle

\section{Introduction}

The study of late type giant variables has been advanced greatly by the use
of infrared photometry in conjunction with large-scale long-term photometric
monitoring. Twenty years ago it was shown that the large-amplitude variables
(Miras) in the Large Magellanic Cloud obey a period-luminosity relation
(Glass and Lloyd Evans, \cite{Glass1981}) which is particularly tight in the
$K$--log$P$ plane ($\sigma$ = 0.13) for both O- and C-rich stars (Glass et
al. \cite{Glass87}; Feast et al. \cite{Feast1989a}). During the last few years,
using the database of the MACHO gravitational lensing experiment, Wood et al
(\cite{Wood99}; see also Wood \cite{Wood2000}) discovered that the smaller
amplitude semi-regular variables of the LMC also obey period-magnitude
relations, parallel to that of the Miras. Wood's work and that of Alard et
al. (\cite{Alard2001}) show that the number of semiregular variables in a
given population greatly exceeds that of Miras.

More recently, the nature of the $K_S$--log$P$ relations have been
investigated with increasing thoroughness by combining the DENIS and 2MASS
near-infrared sky surveys with data from the MACHO, OGLE and MOA large-scale
variability projects. In particular, the SIRIUS $JHK_S$ survey of the
Magellanic Clouds, with greater sensitivity and higher resolution than its
predecessors, has been used to extend the search to fainter magnitudes by
Ita and collaborators (2002, 2004), with the result that the distribution of
late-type variables in the $K_S$--log $P$ plane is now seen to be more
complex than realised at first. Ita et al. (\cite{Ita2002}) and Kiss \&
Bedding (\cite{Kiss2003},\cite{Kiss2004}) showed that variability occurs on
the upper red giant branch along sequences that extend those of the AGB to
fainter mags and are slightly offset from them in log $P$. Cioni et al.
(\cite{Cioni2000}) observed that a minimum exists in the $K$ magnitude
distribution of variables in the SMC and this has been interpreted as
relating to the tip of the RGB, which is metallicity dependent at $K$ (e.g.,
Ita et al. \cite{Ita2002}; Kiss \& Bedding, \cite{Kiss2004}).

$K_S$--log$P$ relations are also found among the late-type variables of the
Milky Way galaxy. Glass \& Schultheis (\cite{Glass2003}) have
cross-correlated the DENIS and MACHO surveys of the inner Bulge field
NGC 6522, whose contents are at an approximately uniform distance from the
sun, and have found relations similar to those of Wood (\cite{Wood2000}).

Multi-periodic behaviour is frequently observed among the late-type
variables. Most conspicuously, the lower- amplitude stars sometimes show two
or more periods differing typically by factors of around 1.4, 2 or their
reciprocals. A further group of stars have short periods of a few tens of
days, modulated by periods about ten times longer. Wray, Eyer \& Paczynski
(\cite{Wray2004}) have analysed period-amplitude information, which is
independent of (often unknown) distances and reddenings, for a very large
sample of galactic Bulge variables from the OGLE database and have shown
them to display complex multi-periodic behaviour, similar to that found in
the Magellanic Clouds.

The mass-loss characteristics of late-type giants have been studied with the
aid of data from the ISOCAM camera on the ISO infrared satellite, which has
observed fields in NGC 6522 as part of the ISOGAL survey (Omont et al.
\cite{Omont2003}) and in the Magellanic Clouds as part of the Magellanic Cloud
Mini-Survey (ISO-MCMS; Loup et al. (2004), in preparation; Cioni et al.
\cite{Cioni2001}, \cite{Cioni2003}). Such mass-losing AGB stars are likely
to be the most important contributors of re-cycled material to the
interstellar medium in many galaxies. Cioni et al. (\cite{Cioni2001},
\cite{Cioni2003}) have analysed data from the two Magellanic Clouds as seen
by MACHO, DENIS and ISO-MCMS.

Alard et al. (\cite{Alard2001}), in an analysis of stars detected at 7 or 15
$\mu$m by ISOGAL in the NGC 6522 field (Glass et al. \cite{Glass99a}), showed
that they are all regular or semi-regular variables and that significant
mass-loss is frequently found in red variables with periods of 70d or
longer; i.e., it is not confined to Mira (by which we mean large-amplitude)
variables. Using a complete spectroscopic survey of an overlapping field by
Blanco (\cite{Blanco86}), it was possible to show that the variable giants
belong to the later M spectral types and that the mass-losing ones in
particular belong to the latest types (Glass \& Schultheis
\cite{Glass2002}).

The aim of the present study has been to compare three fields in the Galaxy
and the two Magellanic Clouds, as they appear in the 2MASS, MACHO and ISO
databases, using the methods previously applied to NGC 6522, with the aim of
seeing how the overall variability, colour, luminosity and mass-loss
properties of late-type giants differ from one to another. Fields containing
similar numbers of stars observed by 2MASS, MACHO and ISO have been
investigated. In an analysis of DENIS data, Lebzelter, Schultheis
\& Melchior (\cite{Lebzelter2002}) have already shown that the $I-J$ colours
of Bulge variables have a larger range ($\sim$ 4 mag) than those in the LMC
($\sim$ 2 mag), which they attribute to the greater spread in metallicity of
the Bulge compared to the LMC.

\section{Sample selections}

All stars with $M_{K_{S,0}}$ $<$ -- 4.75 were extracted from the 2MASS
Catalog for three well-defined fields in the galactic bulge and the two
Magellanic Clouds. This corresponds to about $K_S$ = 9.75 in NGC 6522, 13.5
in the LMC and 13.95 in the SMC. The sizes of the fields were chosen to
obtain comparable numbers of objects. Their limits are shown in Table 1. The
ISO fields are considerably smaller.

\begin{table}
\caption{2MASS fields in NGC 6522, LMC and SMC}
\begin{tabular}{cccc}

field & NGC 6522 & LMC & SMC\\
\hline
$\alpha_{min}$[deg]& 270.5796&80.9309&13.6301\\
$\alpha_{max}$[deg] &271.2432&81.7392&15.6689\\
$\delta_{min}$[deg] &--30.2773&--69.9094&--73.3379\\
$\delta_{max}$[deg]  &--29.7788&--69.6405&--72.9129\\
Number& 1782 & 1809 & 1649\\
 $\rm K_{lim}$[mag]& 9.75 & 13.50 & 13.95\\
Area[arcmin$^2$]& 1032 & 271 & 905 \\
\hline
\end{tabular}
\end{table}

We used the same field size for NGC 6522 as described in Alard et al.
(\cite{Alard2001}) but taking the 2MASS catalog instead of DENIS. The total
number of 2MASS sources is 1782 using a $K_S$ limit of 9.75 (see Table 1). 
For the LMC and SMC, we have chosen fields where ISOCAM observations
are available (Cioni et el. \cite{Cioni2003}). Using the final version of
the 2MASS catalogue, we obtain for the SMC 1649 sources and for the LMC 1809
sources, respectively. The cut-off limits in $K_S$ are given in Table
1.

\section{Extraction of the light curves from MACHO}

Counterparts for each 2MASS source in the NGC 6522, LMC and SMC fields were
searched for in the MACHO database (http://www.macho.mcmaster.ca) within a
typical radius of $\sim$3$''$. Some 133 sources in the NGC 6522 field do not
show a MACHO counterpart, while the comparable numbers are 91 in the LMC and
53 in the SMC. Occasionally, several MACHO sources were found within 2
arcsec of a 2MASS source. A unique counterpart could usually be selected on
the basis of MACHO $r$ and $b$ value, colour, variability and proximity.
Occasionally, a 2MASS source may be the sum of two or more faint
non-variable MACHO sources.

Many of the missing counterparts occur because each MACHO field has one or
two dead zones, corresponding to gaps between the detector chips, where
cross-identifications could not be made. However, there are also cases where
the MACHO tables have blank `patches', i.e., no sources appear within 3
arcsec of the position being searched. There were 12 such positions in the
LMC and 13 in the SMC. Many of the corresponding 2MASS sources were bright
or `blue' in their $JHK_S$ colours. By means of overlays, they were found to
correspond to very bright stars on the UK Schmidt IR survey plates. In many
other cases, a MACHO counterpart could be found, but the data were either
sparse or saturated. From their positions in the colour-colour diagrams, it
is clear that sparseness of data was usually caused by saturation also,
though a few of them fall among the red giants. The sparseness of the data
for the latter may have resulted from crowding effects. The sparse sources
are included in the near-IR colour-colour and colour-magnitude diagrams
only.

\section{Extraction of period information}

As previously, the red and blue light curves and their frequency power
spectra were plotted in batches of 20. The frequency range searched was from
0 to 0.2 cycles per day (5d to infinity in period) and the frequencies
corresponding to the three most conspicuous maxima in each spectrum were
derived automatically. Each light curve and power spectrum was further
examined by eye. The MACHO flags were disregarded as they were often found
to be too conservative. Instead, poor quality data were revealed by
the light curves and the power spectra. Some data included artefacts which
were noted during the visual inspection. Because of the seasonal nature of
the sampling, the peaks in the power spectra were usually broader in NGC 6522
than in the Magellanic Clouds.

Multiple periodicity was found to be common in all three fields (see also
Wray, Eyer \& Paczynski \cite{Wray2004}). We again made a distinction
between doubly-periodic variables, which show a long period (usually several
hundred days) superimposed on a short period (usually several tens of days)
and stars that are multi-periodic with comparable periods (usually several
tens of days).

The Fourier spectra of the doubly-periodic stars were sometimes dominated by
harmonics of the (non-sinusoidal) longer period variations, so that their
short periods were not always among the three highest peaks that were
extracted automatically. Occasionally also there was poor agreement between
the blue and red data. In the first-mentioned category, the short period
could usually be found among the lesser peaks of the Fourier spectrograms
and, in the more difficult cases, a visual estimate of period could be made
from coherent portions of the light curves. The short and long periods were
recorded together with visual estimates of their long-term average
peak-to-peak amplitudes.

The stars with secondary short periods often showed more than two of them.
We recorded the two most prominent short periods and the long period, if
present, as well as the full amplitude of the short and long-period
variations, averaged by eye from the light curves. The ratios of the short
periods were plotted against the log of the dominant period as in fig.\ 9 of
Glass \& Schultheis (\cite{Glass2003}) for each field. The frequency of
detectability of secondary short periods was lower in the Magellanic Clouds
than in NGC 6522 but there were no obvious differences in the general
character of the plots.

The Mira variable D9 in the NGC 6522 field was too bright for MACHO and has
been assigned the approximate period 400d (Lloyd Evans\cite{Lloyd1976}) and
an arbitrary $r$ amplitude of 4 mag for completeness.
  
The few Cepheid variables present in the Magellanic Cloud data were apparent
from their MACHO light curves. Only one eclipsing binary, possibly of
contact type, MACHO 211.16650.11, was found, lying in the direction of the
SMC.

\section{Amplitudes}

The full amplitude of the short (and long, if present) period variations
were recorded from eye estimates, using the MACHO $r$ band data or, if
lacking, the $b$. In the case of multiple short periods, the overall
peak-to-peak amplitude was recorded. For doubly periodic variables, both
amplitudes were noted. Generally, amplitudes were categorized as 0, 0.05,
0.1, 0.15, 0.2, 0.3, 0.4 ... mag. The 0.05 category includes stars whose
variability was just detectable. It is estimated that variability with
peak-to-peak amplitude of $\geq$0.03 mag was detected.

\section{Discussion: Near-IR}

In what follows, we have assumed a distance modulus for the NGC 6522 field of
14.7, for the LMC 18.5 and for the SMC 18.94. The extinction values that we
took were: $E_{B-V}$ = 0.5 for NGC 6522, $E_{B-V}$ = 0.15 for the LMC and
$E_{B-V}$ = 0.065 for the SMC. These were applied to the $JHK_S$ data
according to the extinction law given by Glass (\cite{Glass99}). The plots
given in this paper make use of the de-reddened absolute magnitudes for ease
of comparison. It should be noted that the distance modulus of 14.7 used for
the NGC 6522 field (Glass et al. \cite{Glass95}) is higher than that commonly
accepted. It is retained here because it gives good fits to the $M_{K_S}$,
log$P$ sequences defined from the Magellanic Cloud samples.

The stars plotted in the $JHK$-region diagrams are divided into the
following classes: crosses: saturated and sparse stars; circles:
non-variables; solid dots: small- amplitude variables; boxes:
doubly-periodic stars; asterisks: large-amplitude variables. Note that
doubly- periodic stars were only counted as such if the full MACHO $r$
amplitude of the longer period exceeded 0.2 mag. Large-amplitude variables
are those with full MACHO $r$ amplitude $>$ 1.0 mag.

\subsection{Colour-magnitude diagrams}

\begin{figure*}
\begin{minipage}{17.5cm}
\epsfxsize=16cm
\epsffile[28 62 539 786]{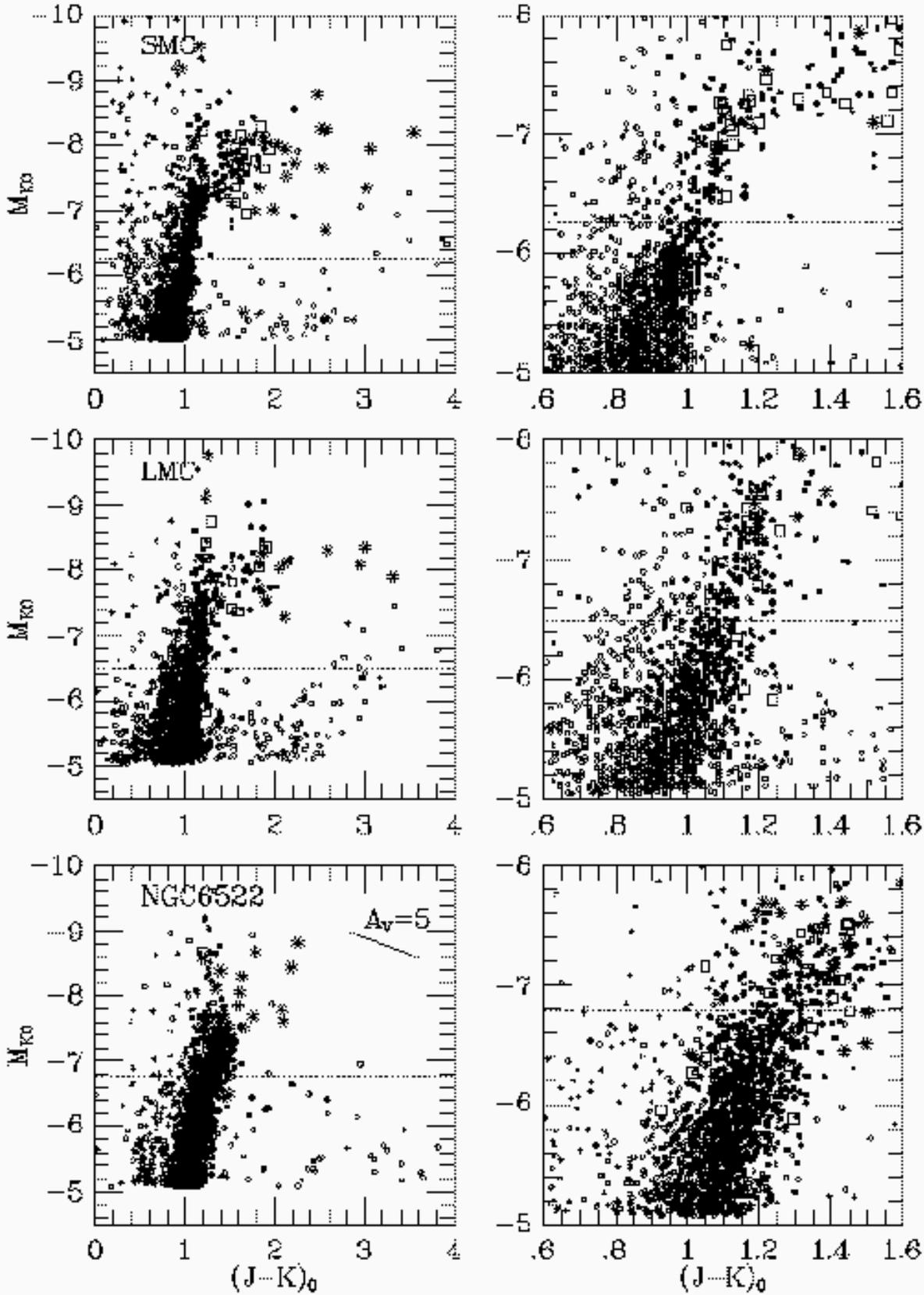}

\caption{{\it Left:} $M_{K_{S,0}}$ vs. $(J-K)_{0}$ diagram for the
NGC 6522 field (bottom), the LMC (middle) and the SMC (top). Saturated and
sparse stars are indicated as crosses, non-variable stars as
open circles, semiregular variables as filled circles, doubly-periodic SRVs
as open squares and large amplitude Mira variables as asterisks. {\it
Right:} Enlargement of densest portion of diagrams.  Most of the points
towards the bottom right of the distributions, i.e., with
$(J-K_S)_0$ $>$ 1.2 -- 1.4 and $M_{K_{S,0}}$ $>$ -- 7, are probably the
result of photometric errors or crowding, since they are constant or of
small amplitude (and are thus not large amplitude variables surrounded by
dust shells). The dotted lines represent the estimated levels of the RGB
tips.}

\label{CMD}

\end{minipage}
\end{figure*}

The left side of Figure~\ref{CMD} shows the colour-magnitude diagrams of the
sources detected in MACHO and 2MASS for the NGC 6522 field, the LMC and the
SMC. The densest part of each diagram is enlarged on the right.

\begin{figure}
\epsfxsize=8cm
\epsffile[28 423 539 786]{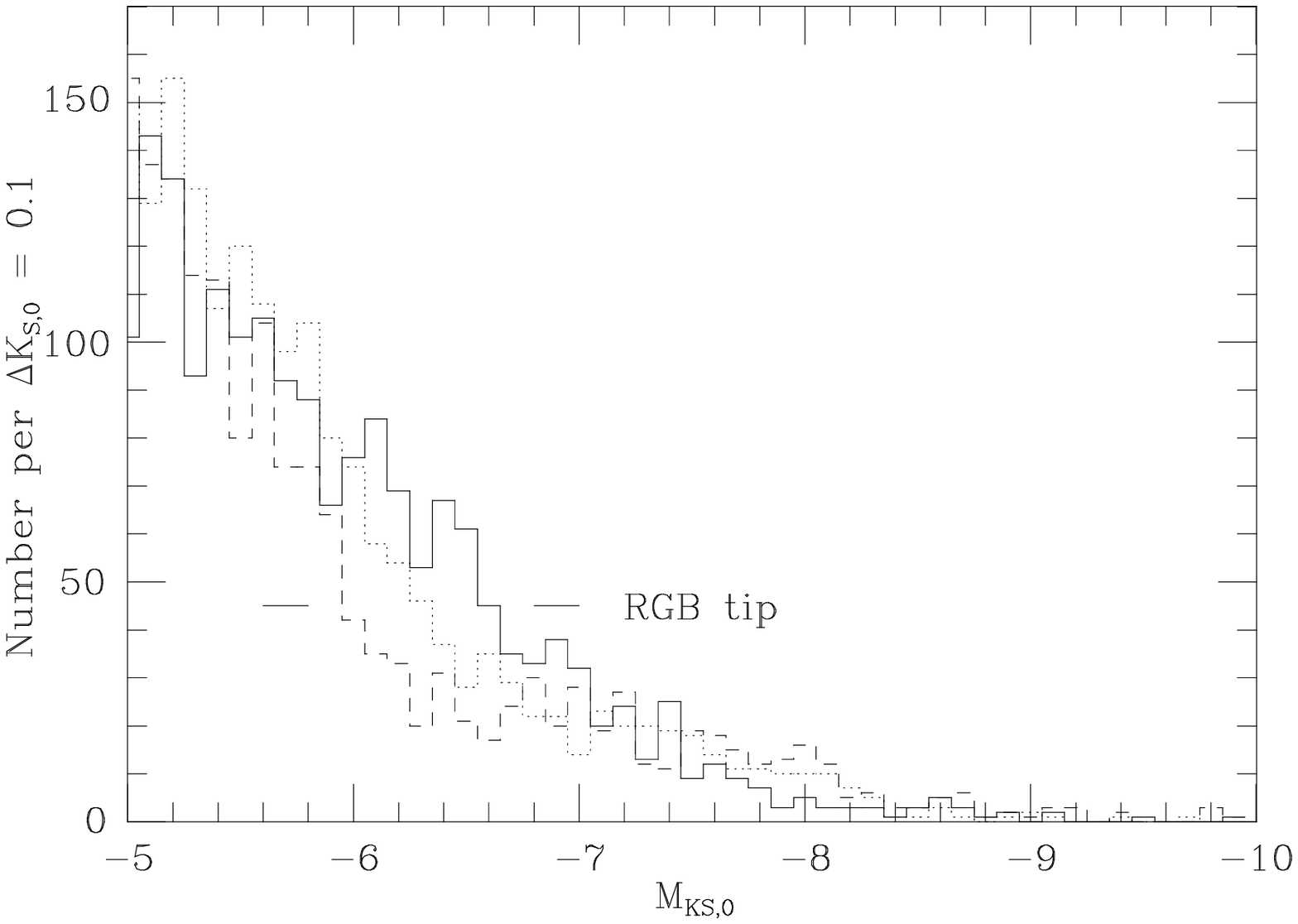}
\caption{Comparative histogram of $K_{S,0}$ for the three fields to show the
sudden drop-off at the RGB tip. The NGC 6522 field is given by the solid
line, the LMC field by the dotted line and the SMC field by the dashed line.}
\label{histo.ps}
\end{figure}

All three fields are dominated by the giant sequence, composed of the AGB
(mainly upper part) and the RGB (lower part). The most conspicuous trend is
that the giant branches move blueward in $(J-K_S)_{0}$ colour as one
progresses from the NGC 6522 to the SMC fields. A lower metallicity is known
to lead to a shift of the giants towards higher temperatures in the HR
diagram and thus to bluer colours. According to Kiss \&
Bedding(\cite{Kiss2004}), the tip of the RGB occurs at $M_{K_{S,0}}$ = --
6.48 in the LMC and -- 6.26 in the SMC. Our
$K$ histograms (fig. \ref{histo.ps}) show that the steep fall-off in numbers
which occurs at this point (see also Cioni \cite{Cioni2000}) is separated by
about 0.3 mag between the SMC and the LMC and by a similar amount between
the LMC and the NGC 6522 field. The level of the RGB tip is shown as a
dotted line in each case (fig. ~\ref{CMD}). The fall-off in the number of
stars is accompanied by a noticeable decrease in average $(J-K_S)_0$ colour
in the LMC and SMC.

According to the work of Ferraro et al. (\cite{Ferraro2000}) on
galactic globular clusters,
\[ 
M_{K_{0}}^{\rm Tip} = -(0.59 \pm 0.11) [\rm Fe/H] - (6.97 \pm 0.15).
\]
There is a small transformation between $K_S^{2MASS}$ and $K^{SAAO}$, so
that, if we take the Kiss \& Bedding figures and assume that the NGC 6522
field RGB tip is 0.3 mag brighter than that of the LMC, we obtain rather low
values of the metallicities for the NGC 6522, LMC, and SMC fields of [Fe/H]
= -- 0.32, \mbox{-- 0.85} and -- 1.22 respectively.

The average $(J-K)_0$ value at a given $M_K$ is also an indicator of
metallicity. We have determined this to be (for -- 5.25 $>$ $M_{K_{S,0}}$ $>$
-- 5.75) 1.09, 0.99 and 0.89 for the NGC 6522, LMC and SMC fields
respectively. To place these values on the SAAO system, for which the
Ferraro et al. (\cite{Ferraro2000}) relation is valid, the transformation
given on the 2MASS website must be applied
\begin{eqnarray} \nonumber 
(J-K_S)_{2MASS} = (0.944 \pm 0.012)(J-K)_{SAAO} \\ - (0.005 \pm 0.006)
\lefteqn \nonumber
\end{eqnarray}
The $(J-K)_0$ values then become 1.16, 1.05 and 0.95 on the SAAO system.
Ferraro et al. (\cite{Ferraro2000}) give
\[
(J-K)_0^{-5.5} = (0.21 \pm 0.01)[\rm Fe/H] + (1.13 \pm 0.02)
\]
The metallicities according to this criterion are then 0.14, -- 0.38 and
-- 0.86 for the NGC 6522, LMC and SMC fields, respectively. Terndrup, Frogel
\& Whitford (\cite{terndrup1990}) found [M/H] $\sim$0.2 from M giants in
inner Bulge fields. Young ($<$ $10^9$ yr) populations in the Large
Magellanic Cloud have [Fe/H] $\sim$ -- 0.2 and in the Small Cloud $\sim$ --
0.5, but for older populations these figures can be as much as 1 dex smaller
(Feast, \cite{Feast1989b}). The metallicity values derived from the RGB tip
magnitudes are lower by about 0.4. A discussion of the uncertainties in
these relations is given by Ferraro et al. (\cite{Ferraro2000}). The present
results, though not consistent in absolute metallicity, at least agree that
metallicities differ by about 0.4 to 0.5 dex between the Bulge and the LMC
and between the LMC and SMC fields. It can also be that age affects the
stars at the tip of the RGB differently from those at $M_K$ = -- 5.5.

\subsubsection{Relation to theoretical isochrones}

Differences in tip RGB $K$ luminosities due to age and metallicity can be
estimated using theoretical isochrones from Girardi et al.
(\cite{Girardi2000}). Consider first the differences in $M_{K_S}^{\rm Tip}$
for two populations of {\it fixed metallicity} that are about 3 and 10 Gyr
old. These amount to 0.16, 0.23 and 0.27 mag for $Z$ = 0.004, 0.008, and
0.019, respectively, approximately corresponding to the average
metallicities of the SMC, LMC and NGC6522. Clearly, differences due to age
are stronger in a metal-rich population.

If we compare $M_{K_S}^{\rm Tip}$ for populations with {\it fixed ages} of
10\,Gyr but differing metallicities we find the difference in $M_{K_S}^{\rm
Tip}$ between Z=0.004 and Z=0.008 is 0.32 mag and the difference between
Z=0.008 and Z=0.019 is 0.19 mag. For 3\,Gyr old populations, these figures
become 0.25 and 0.15 mags. Differences due to metallicity are thus somewhat
greater for an older population.

It is believed that each of these galaxies contains a composite stellar
population and therefore the observed 0.3 mag difference in the tip of the
RGB is most probably the result of both age and metallicity effects. Since
the Clouds are closely related and it is possible that their RGB populations
have the same age, or the same mixture of ages, may be that the $\geq$0.3
mag difference can be explained by a metallicity effect alone. However, more
realistically, within each Magellanic Cloud a spread in metallicity of order
0.75 dex is likely to be present (Cioni \& Habing \cite{Cioni2003a}) or at
least to have been present at the time their AGB stars were formed.


\subsubsection{Further comments on the colour-magnitude diagrams}

There are increasing numbers of stars redward of the top of the AGB at
$M_{K_{S,0}}$ $\sim$ --8 as lower metallicities are approached. The reddest of
these stars in the NGC 6522 field are O-rich Mira variables, whereas in the
Magellanic Clouds they may also be C-rich Miras. However, it is clear that
there are many low-amplitude and double-period Magellanic Cloud variables
with elevated $(J-K_S)_0$ that have no counterparts in the NGC 6522 field.
These are probably C-rich. It is well-known (Lloyd Evans
\cite{lloydevans1988}) that carbon stars are more abundant among large
amplitude variables in the SMC than the LMC.

It is also curious that there are no doubly-periodic variables fainter than
the red giant tip in the SMC field, whereas there are three in the LMC and
many in NGC 6522.

There are considerable populations of non-variable stars around 0.3 mag
blueward of the giants in the Magellanic Cloud fields, though not in
NGC 6522. Many of these have $(J-K_S)_{0} < 0.5$ and are most likely
galactic foreground objects. There is a higher proportion of galactic
foreground stars in the LMC ($(J-K_S)_{0} \sim 0.7$), than in the SMC.
(see Nikolaev \& Weinberg \cite{Nikolaev2000}).

\subsection{Colour-colour diagrams}

\begin{figure}
\epsfxsize=8cm
\epsffile[28 -9 443 786]{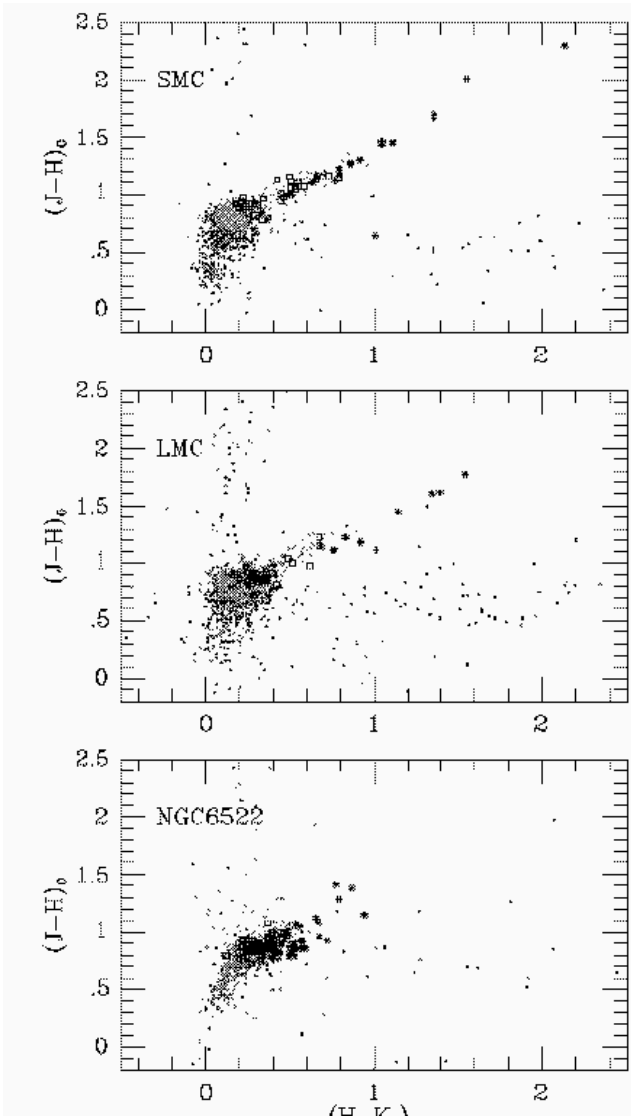}

\caption{$\rm (H-K_S)_{0}$ vs. $\rm (J-H)_{0}$ colour-colour diagram for the
NGC 6522 field (bottom), the LMC (middle) and the SMC (upper). Non-variable
and saturated stars are indicated by black dots, semiregular variables by
grey, doubly--periodic SRVs by open squares and Mira variables by
asterisks. Some features can be seen more clearly on a colour version of
this figure, available at ...}

\label{CCD}
\end{figure}

Figure ~\ref{CCD} shows the $(J-H)_{0}$ vs. $(H-K_S)_{0}$
colour-colour diagram of the MACHO variables.

Especially the LMC diagram contains a large number of spurious points to the
right of the concentrated area and some others directly above it. These are
suspected to arise from errors in the 2MASS photometry and errors in the
cross-identifications between bands. Similar effects occur in the DENIS
data. This diagram can be seen in colour on the ... web site, with constant
stars coded blue, saturated stars magenta, small-amplitude variables green,
doubly-periodic cyan and Miras red.  The constant stars (many of them in the
foreground) extend the concentrated area in the direction of bluer and less
luminous stars (see the colour version for a clearer picture).

While the colour-colour diagrams of the LMC and SMC look rather similar, the
stars in the NGC 6522 field have a smaller overall range in $(J-H)_{0}$ and
$(H-K_S)_{0}$ because of the red colour of the many carbon stars in the LMC
and SMC which do not occur in the Bulge. The $J-H$ and $H-K$ colours are
sensitive to metallicity and atmospheric extension (Bessell et al.
\cite{Bessell96}).

The doubly-periodic stars in the Magellanic Clouds are among the reddest of
the small-amplitude variables (even their longer periods rarely exceed 1.0
mag). They can be redder than their counterparts in NGC 6522. At least some
of them will be carbon stars. However, Groenewegen (\cite{groen2004}) has
cautioned against deciding on carbon- vs oxygen-richness on near-IR colours
alone.

\subsection{Magnitude distributions}

Stars whose peak-to-peak amplitudes were lower than 0.03 mag in $r$ were
generally not detectable as variables. This limit is somewhat subjective and
depends on the quality and quantity of the data for the individual objects,
neither of which are absolutely uniform. The possibility exists that some of
the objects classified as non-variable may, in fact, be variables of very
small amplitude. The dependence of variability on spectral type was
discussed by Glass \& Schultheis (\cite{Glass2002}).

\begin{figure}
\epsfxsize=8cm
\epsffile[28 -9 476 786]{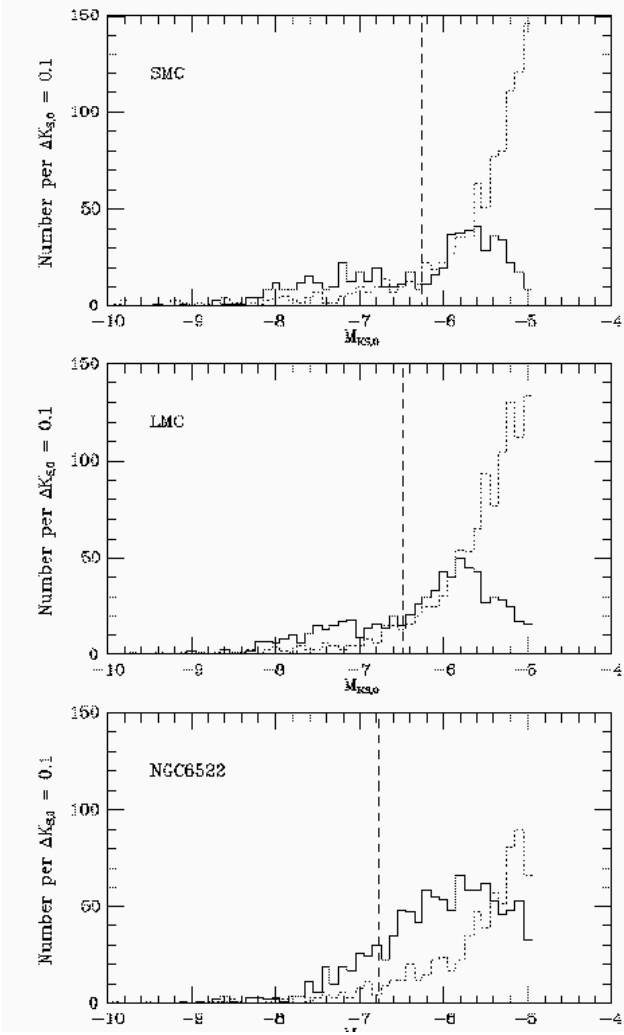}

\caption{Histogram of $\rm K_{S,0}$ for variable stars (full line) and
non-variable stars (dotted line). The dashed lines indicate
approximately the locations of the tips of the RGBs (see text).}

\label{hist1}
\end{figure}

Figure \ref{hist1} shows the distributions of absolute $K_{S,0}$ magnitude
for the variable and non-variable stars. Indicated are also the locations of
the tips of the RGB as discussed previously.  It is clear that the variable
stars on the AGB dominate the non-variables by far. While this is also true
for the SMC and LMC, the ratio of variables to non-variables is much higher
in the NGC 6522 field than in the LMC and the SMC. This tendency remains
dominant for the Bulge up to about $M_{K_{S,0}} \sim -5$ (see
Fig.~\ref{hist1}) which is presumed to be the completeness limit for
detecting variable stars. For the LMC and SMC the characteristics of 2MASS
cause this limit to be about $M_{K_{S,0}} \sim -6$. Thus the proportion of
variable stars beyond the tip of the RGB decreases as we go downwards in
metallicity from the NGC 6522 field to the SMC field.

A further important characteristic is that the proportion of luminous
AGB stars with $M_{K_{S,0}}$ $<$ --7.5 increases with decreasing metallicity.
It will be seen later (section \ref{Mid-IR}) that these constitute most of
the stars detected by ISO in the Magellanic Clouds.

\subsection{Period-$M_{K_{S,0}}$ relations}

\begin{figure}
\epsfxsize=8.2cm
\centerline {\epsfbox[28 -9 445 786]{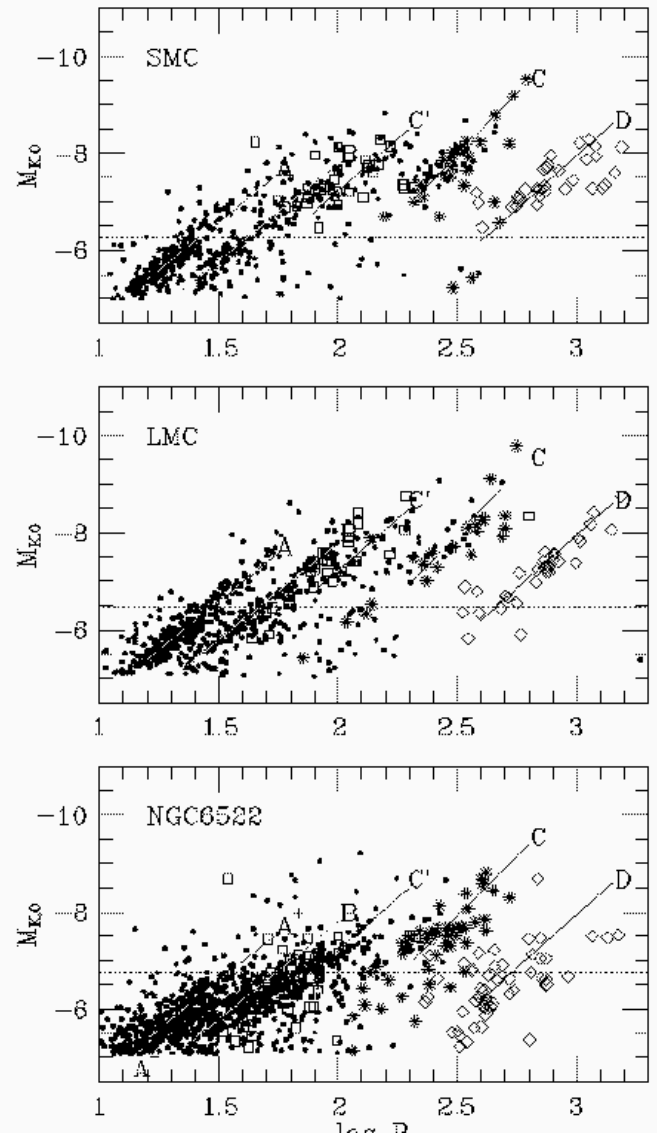}}

\caption{$\rm log\,P$ vs. $M_{K_{S,0}}$ relation for the NGC 6522 field
(bottom), the LMC (middle) and the SMC (top). Small amplitude variables are
indicated by filled circles, doubly-periodic SRVs by open squares (short
periods) and diamonds (long periods) and Mira variables by asterisks. The
straight lines are eye fits to the LMC sequences. The level of the top of
the RGB is indicated by a dotted line.}

\label{PL}
\end{figure}

Figure  \ref{PL} shows the $M_{K_{S,0}}$, log $P$ diagram for the three
fields. The Magellanic Cloud fields show a number of clearly separated
sequences, but in the NGC 6522 field they overlap considerably because of the
depth of the galactic bulge along the line-of-sight.

Diagrams of this kind with much larger numbers of stars have been produced
for the Magellanic Clouds by other authors (e.g., Ita et al. \cite{Ita2004},
Kiss \& Bedding \cite{Kiss2003}, \cite{Kiss2004}) and some of the subtler
aspects of the distributions are better seen in their work.

The scatter about relation C (the Miras) is higher than in previous work
(e.g., Glass et al. \cite{Glass95}) because these variables have large
amplitudes and only single-epoch observations at $K_S$ are used here.

The LMC diagram has been fitted by eye with a number of lines that indicate
the positions and extents of the discernable sequences. These lines have
been copied onto the other fields to aid in intercomparisons. Ita et al.
(\cite{Ita2004}) (Fig 1) gives outline boxes for the sequences in the two
Magellanic Clouds. To the original ABCD of Wood et al. (\cite{Wood2000}) they
have added a sequence C$'$, which we see clearly in both these fields, though
they do not separate it from their B$^+$ in the SMC. They further divide
Wood's A and B sequences at the tip of the RGB into separate ones (A$^-$,
A$^+$, B$^-$, B$^+$). We have divided only the A sequence in our diagram.

The slopes of the sequences seem to be very similar between the two
Magellanic Clouds and are probably also compatible with the NGC 6522 field,
though here the scatter is too large for a strict comparison to be made.

We have indicated by dotted lines the approximate locations of the RGB tips
as discussed previously.

The short periods of the double-period variables appear on the B and C$'$
sequences, but not on A or C. Their long periods appear along a sequence of
their own (D). The nature of the long-period variations is not yet
understood.

There are strong differences in this diagram between the galactic and
Magellanic Cloud samples. In general, the Magellanic Cloud sequences extend
to brighter absolute magnitudes than the NGC 6522 field and have higher
proportions of luminous variables. Although this conclusion may not be
statistically firm for the C (Mira) sequence, it is very clear in the A, B
and C$'$. In fact, the A sequence in NGC 6522 may not even extend above the
RGB limit. Kiss \& Bedding (\cite{Kiss2003}) show that stars on these
extensions frequently have red $J-K_S$ colours, which makes them likely to
be C-rich.

At lower metallicities, the lower end of the C sequence, normally associated
with Miras, is populated by increasing numbers of small-amplitude variables.
These are numerous in the SMC and very rare in NGC 6522.

\begin{figure}
\centering
\epsfxsize=8cm
\epsffile[20 150 580 700]{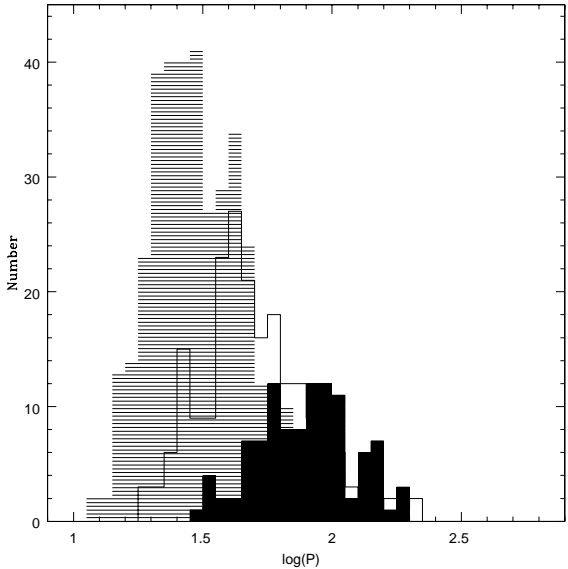}

\caption{Period distribution of small-amplitude variable stars with $\rm
0.05 < A < 0.2$ for NGC6522 (shaded histogram), the LMC (empty histogram)
and the SMC (filled histogram). }

\label{low.ps}
\end{figure}

\begin{figure}
\epsfxsize=8cm
\epsffile[20 150 580 700]{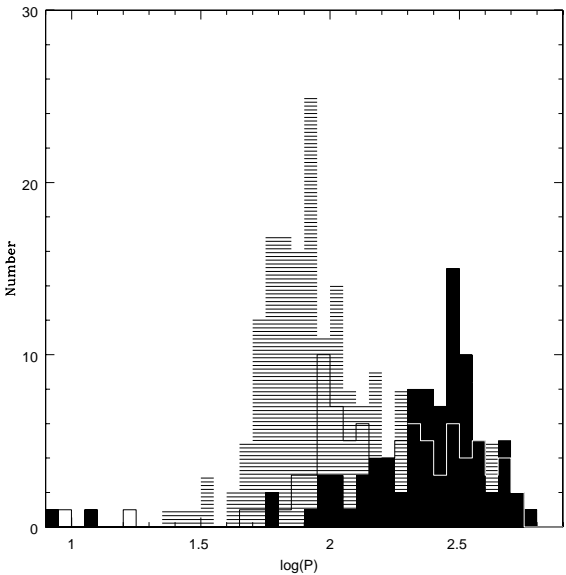}

\caption{Period distribution of large-amplitude variable stars with $\rm A >
0.2$ for NGC6522 (shaded histograms), LMC (empty histogram) and the SMC
(filled histogram). }

\label{high.ps}
\end{figure}

\subsection{Amplitude distributions}

Figure \ref{low.ps} shows the period distribution for the small-amplitude
($\rm < 0.2\,mag$) and Fig.~\ref{high.ps} that for the large-amplitude
variables in the three different fields. They show that the amplitudes (both
small and large) decrease progressively from the NGC 6522 to the SMC fields.
A longer period is reached at a given amplitude as the metallicity
decreases, or alternatively a given amplitude is attained at a longer period
when the metallicity is low. For example, stars with amplitude smaller than
0.2\,mag show their peak at log $P$ $\sim$ 1.35 in NGC 6522, at log $P$
$\sim$ 1.6 in the LMC and at log $P$ $\sim$ 1.9 in the SMC.

\begin{figure}
\epsfxsize=8cm
\epsffile[-28 -9 477 786]{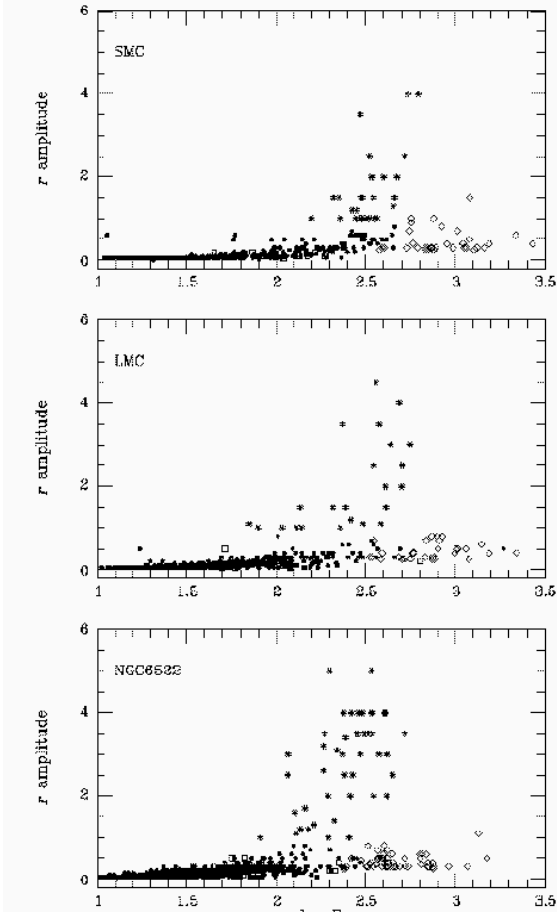}

\caption{General MACHO log$P$ vs. amplitude diagram for the NGC 6522 field
(bottom), the LMC (middle) and the SMC (upper). Non-variable stars are
indicated by crosses, semiregular variables by filled circles,
doubly-periodic SRVs by open squares and Mira variables by asterisks.
Large-amplitude variation occurs at a shorter period in the more metal-rich
NGC 6522.}

\label{AvsP1.ps}
\end{figure}

The general distribution of larger-amplitude variables seems to be very
similar in the Bulge, LMC and SMC. For log $P$ up to
$\sim$2.0 in the galaxy and $\sim$2.3 in the Clouds, the amplitudes stay
small ($\rm < 0.2\,mag$) before increasing dramatically to 1 $< A <$ 5 mag
(see Fig. \ref{AvsP1.ps}).  However, the fraction of all (including the
large-amplitude) variables is smaller for metal-poorer environments. Indeed,
one expects that the visual amplitudes will in general be smaller for lower
metallicity which makes the bands of highly sensitive molecules like TiO
(Reid \& Goldston \cite{Reid2002}) weaker. This would favour small amplitude
variability in metal poor environments and explain the smaller fraction of
large amplitude objects in the SMC compared to the LMC and the galactic
Bulge. This picture is also consistent with the complete lack of Miras in
metal-poor globular clusters (Frogel \& Whitelock \cite{Frogel98}). It is
also the case that stars traditionally classified as carbon Miras in the
Large Magellanic Cloud tend to have smaller amplitudes than their M-type
counterparts (Glass \& Lloyd Evans \cite{gltle2003}).

\subsection{Colour, log $P$ diagrams}

\begin{figure}
\epsfxsize=8.0cm
\epsffile[28 -9 442 786]{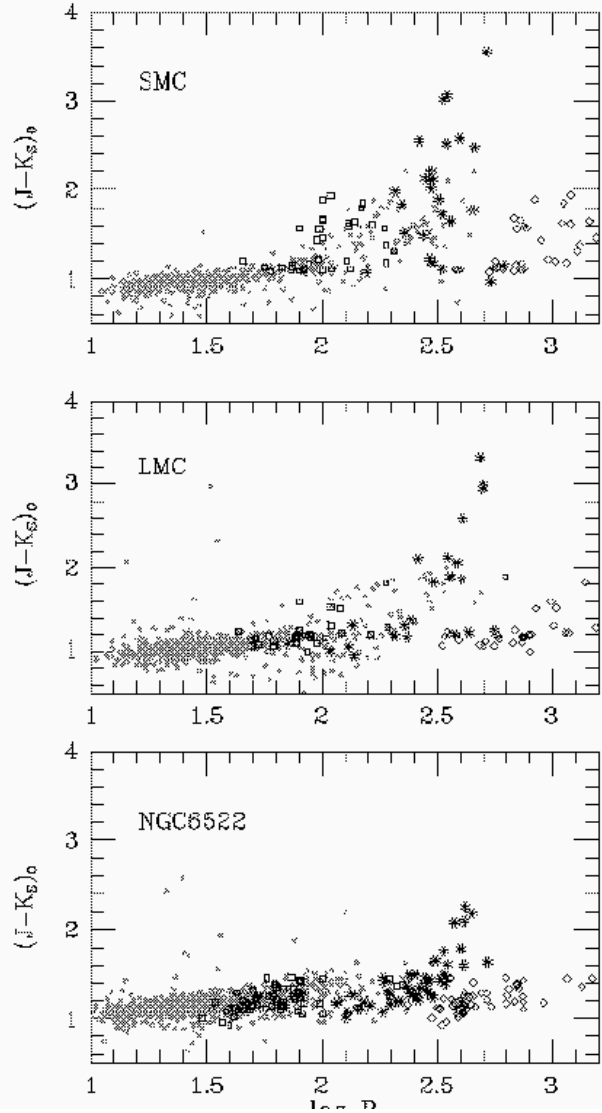}

\caption{$(J-K_S)_0$ vs log $P$ diagrams. The effect of carbon stars is
obvious among the SRVs with periods around 100d and longer as well as the
doubly-periodic stars and the Miras.}

\label{JKvsP.ps}
\end{figure}

Fig. \ref{JKvsP.ps} shows the $(J-K)_0$, log $P$ diagram for each field. The
thickness of the $(J-K)_0$ distribution of the short-period variables (up to
100d) is highest in the NGC 6522 field. The Magellanic Cloud fields show 
increasing numbers of carbon SRVs at log$P > 2.0$. These have
$(J-K_S)$ $\geq$ 1.4 in the Magellanic Clouds; this colour can be used as
a diagnostic tool.  The $(J-K_S)_0$ colour of some Miras is lower than that
of SRVs of similar period, probably because of the onset of water-vapour
absorption which, unlike CO, is contained in the $K_S$ bandpass and thus is
more likely to affect it than the regular $K$ bandpass.

\section{Discussion: Mid-IR}

\label{Mid-IR}

\subsection{ISO observations}

Parts (see Table \ref{tab2}) of our three fields were observed at mid-IR
wavelengths by the ISOCAM camera on ISO.  The fields in the LMC and SMC
formed part of the Magellanic Cloud Mini-Survey (Loup et al. (2004), in
preparation), which used the LW2 (or 7$\mu$, range 5--8.5
$\mu$m, centre wavelength 6.7$\mu$m) and the LW10 (12$\mu$m, range 8--15
$\mu$m, centre 12.0$\mu$m) filters.

The NGC 6522 Baade's Window was observed as a fiducial field during the the
ISOGAL survey of the galactic plane and inner Bulge (Omont et al.
\cite{Omont2003}), using the same 7$\mu$m filter but a different longer
wavelength one, the LW3 (or 15$\mu$m filter, range 12.0--18$\mu$m, centre
14.3 $\mu$m). The following conversions between flux and magnitude are
used:\footnote{For [12] see Origlia et al, ApJ 571, 458, 2002.}

\[ [7]= 12.38-2.5 {\rm log} F_{\nu,LW2} {\rm (mJy)}\]
\[ [12]= 11.13-2.5  {\rm log} F_{\nu,LW10} {\rm (mJy)}\]
\[ [15]=10.79-2.5  {\rm log} F_{\nu,LW3} {\rm (mJy)}\] 

\begin{table}
\caption{ISO fields in NGC 6522, LMC and SMC}
\label{tab2}
\begin{tabular}{ccccc}

field & NGC 6522 & LMC & SMC1 & SMC2\\
\hline
$\alpha_{min}$[deg] & 270.624 & 80.932  & 13.909  & 14.819  \\
$\alpha_{max}$[deg] & 270.961 & 81.718  & 14.755  & 15.615  \\
$\delta_{min}$[deg] & -30.118 & -69.907 & -73.246 & -73.167 \\
$\delta_{max}$[deg] & -29.834 & -69.641 & -73.006 & -72.929 \\
Number$^1$          & 261  & 334 & \multicolumn{2}{c}{227 (total)} \\
Area[arcmin$^2$] & 225 & 261 & \multicolumn{2}{c}{411 (total)} \\
\hline
\end{tabular}

$^1$Number of ISO sources in field with 2MASS counterparts but irrespective
of MACHO cross-identifications.

\end{table}

No reddening corrections have been applied to the ISO data since they are
small and uncertain. The observational errors in the current sample are
typically $\pm$0.14 mag at each wavelength.

It should be remembered that the ISO and 2MASS observations were not
simultaneous so that the combination of near- and mid-IR data, particularly
of the large-amplitude variables, may be affected.

\begin{figure}
\epsfxsize=8cm
\epsffile[20 17 592 779]{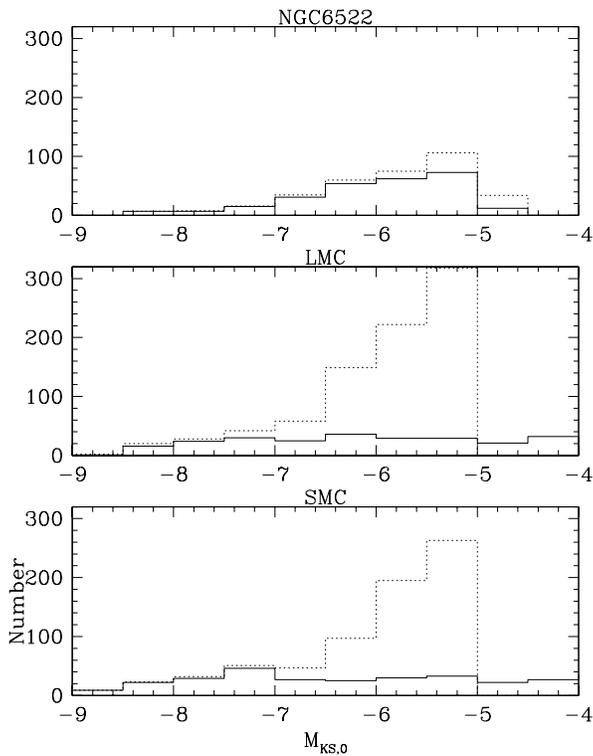}

\caption{Histogram of $M_{K_{S,0}}$ for stars associated with ISO sources
(straight line) and not associated (dotted line). ISO identifications with
2MASS sources are likely to be unreliable fainter than $M_{K_{S,0}}$ = --7
in the Magellanic Clouds and --5 in NC6522.}

\label{hist2}
\end{figure}

The number of ISO detections as a function of the $K$ magnitude is shown in
Fig.~\ref{hist2}. While the ISO detections in the LMC and SMC evidently
become incomplete fainter than $M_{K_0} \sim -7$, the NGC 6522 field can
be regarded as complete down to $M_{K_0} \sim -5.5$. As already pointed out
by Glass \& Schultheis (\cite{Glass2002}), ISO detections are complete for
NGC 6522 at the tip of the RGB.  For the LMC and SMC, only the most luminous
AGB stars are complete. Many of the detected objects have
$(J-K_S)$ $>$ 1.4 (fig. \ref{CMD}) but it will be seen that this does not
necessarily imply that they are  C-rich. The associations less luminous than
$M_{K_0}$ $\sim$ --7 are probably by chance in most cases.  They have been
omitted from further discussion.

\subsection{Near- to mid-IR colour-magnitude diagram}

\begin{figure}
\epsfxsize=8cm
\epsffile[18 -9 469 786]{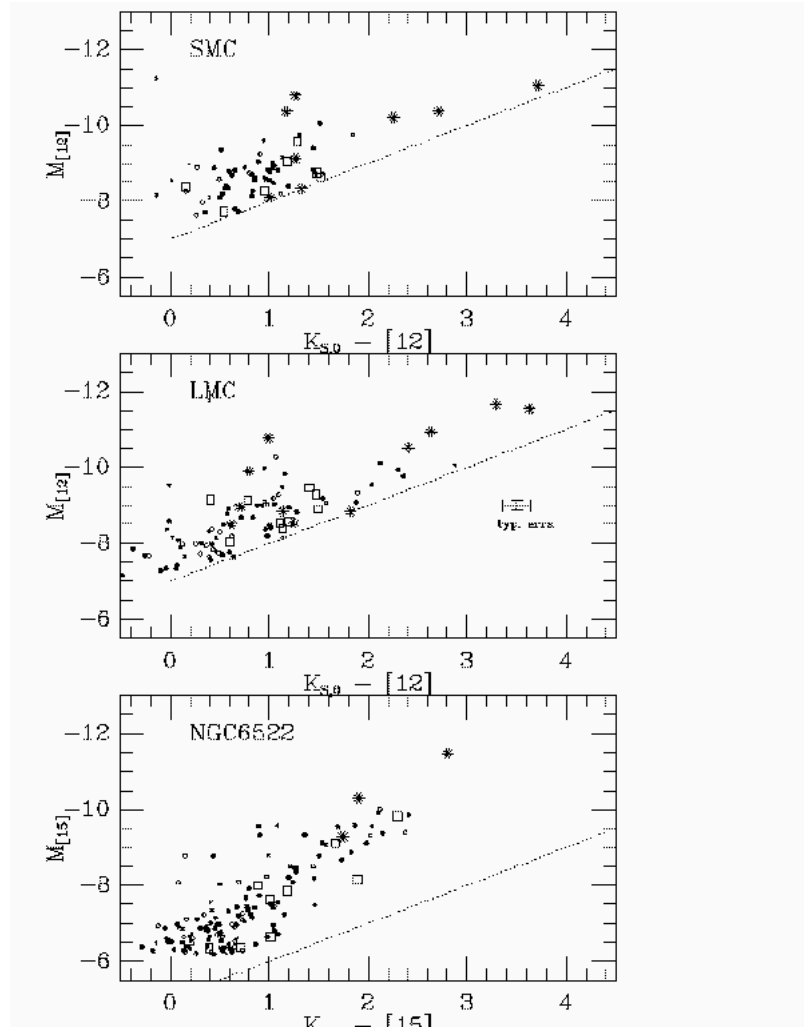}

\caption{[12/15] vs $K$ -- [12/15] colour-magnitude diagram. The dotted line
indicates the likely limit of reliability of the 2MASS-ISO
cross-correlations. Note that objects in NGC 6522 are detected to much lower
luminosities than in the LMC and SMC. However, the Magellanic Clouds have a
substantial population of objects around $M_{12}$ = -- 8 to -- 10 and
$K_{S,0}$ -- [15] = 1 which is not the case in NGC 6522.}

\label{long.ps}
\end{figure}

Figure \ref{long.ps} shows the $K_{S,0}-[12/15]$ vs. [12/15]  diagram. Note
that the LW10 filter which is used for the LMC and SMC is close to the IRAS
12 $\mu$m filter whereas the ISOGAL LW3 filter is centered around
15\,$\mu$m. The LW10 filter includes the 9.7\,$\mu$m silicate and the
11.3\,$\mu$m SiC dust features.  Superimposed is the approximate
completeness limit in $K_{0}$, arising from 2MASS, as described above. Most,
but possibly not all, of the (eliminated) points below this line are
spurious ISO-2MASS associations but some could be real detections affected
by thick circumstellar shells. As already discussed by Alard et al.
(\cite{Alard2001}) and Glass \& Schultheis (\cite{Glass2002}), the $
K_{0}-[15]$ vs. [15] diagram in NGC 6522 reveals a mass-loss sequence in
which the large-amplitude Mira variables are situated at the top of the
diagram. This mass-loss sequence is also shown in Fig.~\ref{long.ps} for NGC
6522.

Cioni et al. (\cite{Cioni2003}) studied long-period variables detected by
ISO in the SMC. They found that Mira variables have large $K_S$ -- LW10
colours, where the carbon-rich objects show on average redder colours than
the oxygen-rich stars.  Figure ~\ref{long.ps} shows that the diagrams for
the LMC and SMC are quite different. The numbers of high mass-losing
semiregular variables with $(K_{S,0} -[12]) > 1.5$ (black dots) seem to
decrease going to lower metallicities and are totally absent in the SMC.
However, one has to be extremely careful concerning selection effects and
small-number statistics.

\subsection{Two near- to mid-IR colour-colour diagrams}

\subsubsection{$(H-K)$ vs $(K-[7])$}

\begin{figure}
\epsfxsize=8cm
\epsffile[28 -9 470 786]{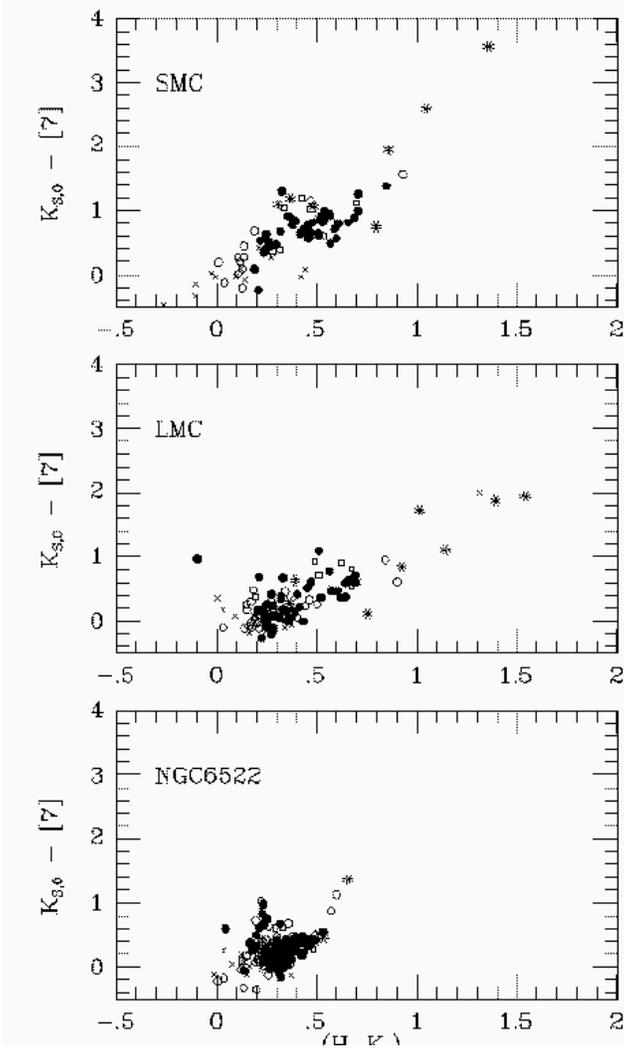}

\caption{$(K_{S,0}-[7])$  vs $(H-K_S)_0$ colour-colour diagrams. Note the
redward extension in both colours mainly caused by the Magellanic Cloud
Miras.}

\label{cc.ps}
\end{figure}

In the near- vs mid-IR colour-colour diagram (fig. \ref{cc.ps}), the range
of colours present in NGC 6522 is small compared to the two Magellanic Cloud
fields. There is a clump of variables around $(H-K_{S})_{0}= 0.3$, $K_{S,0}$
-- [7] = 0.2 in the NGC 6522 and LMC fields. However, this area is blank in
the SMC. This effect is probably due to a lack of ordinary M-type variables
in the SMC sample relative to the other two fields, caused by its lower
metallicity and increased prevalence of C stars. It will be seen later that
very red $(H-K_S)_0$ colours in Mira variables are not necessarily
associated with C-richness.

\subsubsection{$(K-[7)]$ vs $([7]-[12/15])$}

\begin{figure} 
\epsfxsize=8cm 
\epsffile[29 -9 470 786]{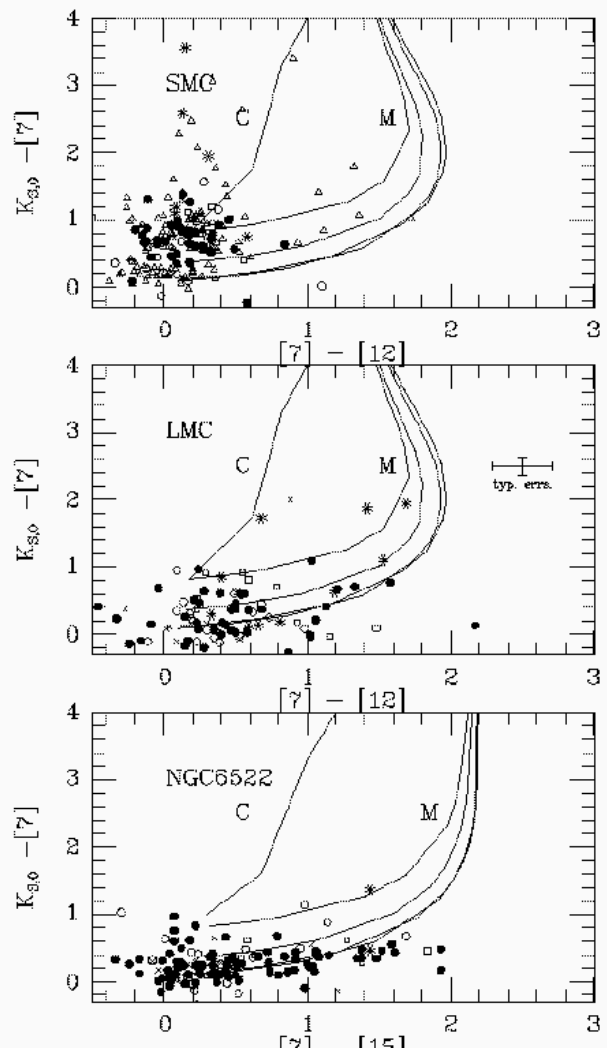}

\caption{$K_{S,0}-[7]$ vs. [7] - [12/15] diagram. Also shown are the colours
of C, M3, M5, M8 and M10 stars with increasing circumstellar shells,
according to the models of Groenewegen (see Loup et al. (2004), in
preparation). The approximate locus of carbon stars is shown by the single
line to the left and the O-rich models are given as four curved lines, with
late types to the right. This diagram distinguishes clearly between (thick)
C- and O-rich circumstellar shells. The SMC data have been augmented with
some extra stars from Cioni et al. (\cite{Cioni2003}) that fall outside the
areas in Table 2 (triangles).}

\label{loup}
\end{figure}

The hybrid $K_{S,0}-[7]$ vs. [7] - [12/15] colour-colour diagram (fig.
\ref{loup}) yields the clearest separation between O-rich and C-rich 
variables. Groenewegen (see Loup et al. (2004), in preparation) has modelled
the development of carbon-rich shells for late-type stars as a function of
their optical thickness. The carbon models use a simple blackbody underlying
spectrum, modified to take into account the 3.1 $\mu$m hydrocarbon feature,
and the dust is composed of SiC and amorphous C. The M-type models (see Loup
et al. (2004), in preparation) combine more realistic central star models
from Fluks et al. (\cite{fluks1994}) with silicate shells.

The LMC Mira-like variables are predominantly of M type and thus O-rich,
although there is one probable C Mira. As expected, there are no stars along
the C locus in the NGC6522 field.

The SMC sample is concentrated towards the blue end of the range in [7] -
[15]. This is mainly the result of the predominance of C-rich stars. The
C-type Mira-like variables stand out from the locus of C-star models but if
$K_S$ mags from DENIS are used, the discrepancy is somewhat reduced. It is
believed that their extreme positions are the result of their variability
and the times of measurement, besides some contributions from the
measurement errors.  In order to reduce the effects of small-number
statistics, data from the larger area described by Cioni et al. (2003) have
also been included.

\subsection{Period-$M_{[7]}$ relation}

\begin{figure}
\epsfxsize=8cm
\epsffile[28 -9 477 786]{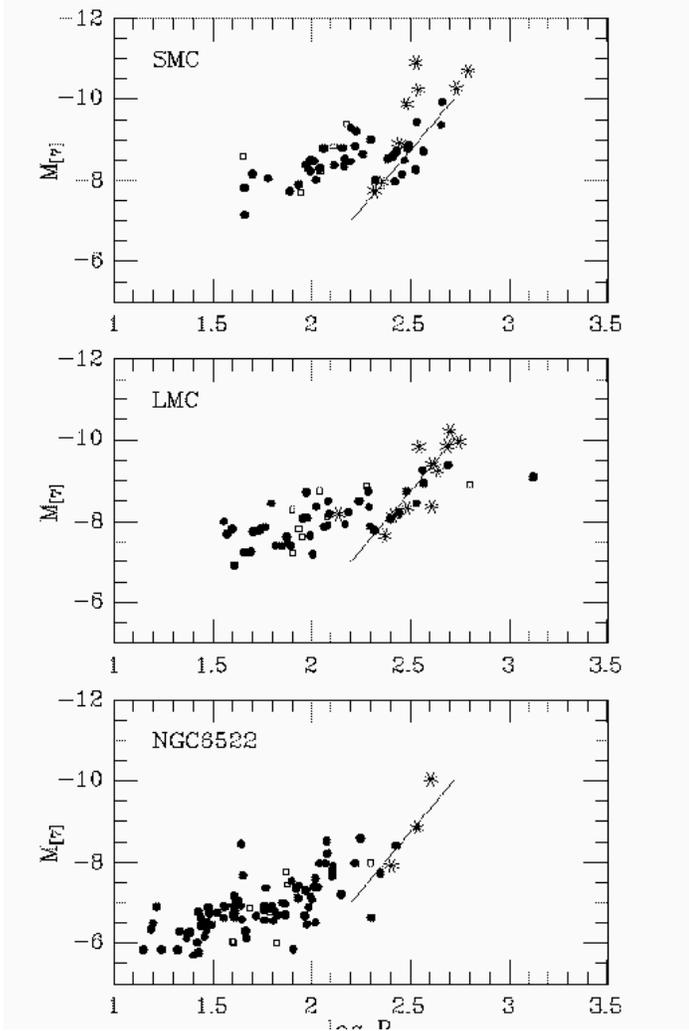}

\caption{Absolute 7$\mu$m magnitude vs log P. A line has been fitted to the
LMC Miras and reproduced at the same position on the other two diagrams.}

\label{7vsP}
\end{figure}

There is a strong suggestion that the period-magnitude relations persist at
7$\mu$m (fig. \ref{7vsP}). Only the Mira relation is really clear because
of the higher photometric errors associated with the ISO measurements.

The stars on the shorter-period sequences disappear as one goes from the
NGC 6522 field to the SMC because of the ISO sensitivity limits.

This diagram shows that the 7$\mu$m zero points of the Magellanic Clouds are
in agreement, though the SMC points could be too high by up to 0.4 mag.

\subsubsection{Mid-IR colour-magnitude diagram}

\begin{figure} 
\epsfxsize=8cm 
\epsffile[28 -9 468 786]{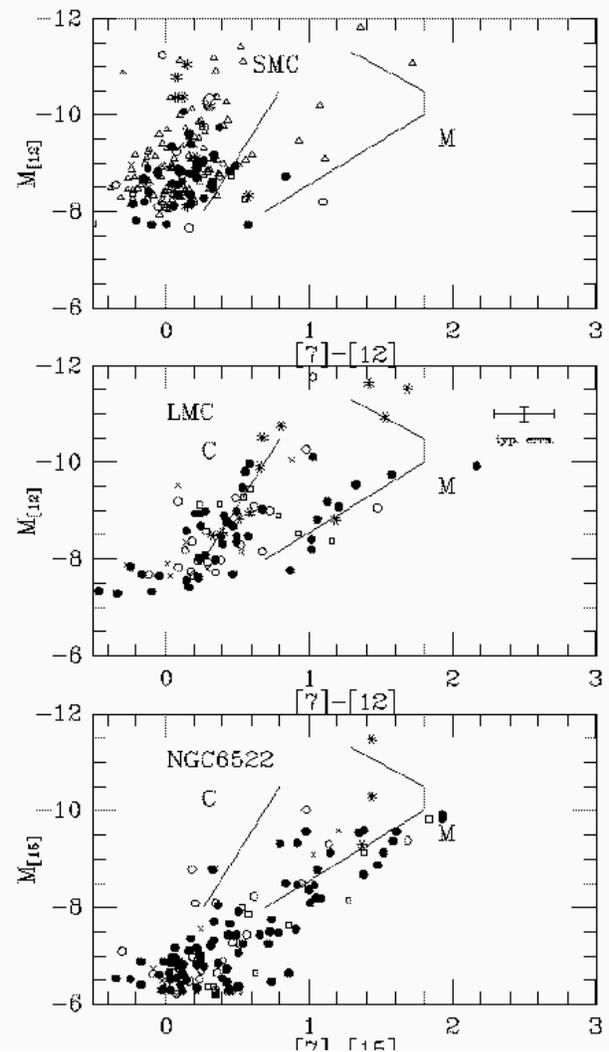}

\caption{$M_{[12/15]}$ vs. [7] - [12/15] diagram. It is again noticeable
that the variables with developed shells are well separated into C-rich and
O-rich sequences. The straight line shows the approximate location of C-rich
objects and the kinked line shows the general trend for M-type objects,
based on the semi-empirical models discussed in the text. The SMC data have
been augmented with some extra stars from Cioni et al. (\cite{Cioni2003})
that fall outside the areas in Table 2 (triangles).}

\label{longcm.ps}
\end{figure}

In the mid-IR colour-magnitude diagram (fig. \ref{longcm.ps}), because of
the different filters, the NGC 6522 and Magellanic field data may not be
strictly comparable. The approximate positions of the semi-empirical models
discussed previously are shown by a straight line for C-rich stars and by a
kinked one for O-rich stars. The shape of the latter line depends only
marginally (in relation to the probable errors) on spectral type but can be
moved vertically depending on luminosity. The fundamental chemical
differences between the three fields are shown well by these diagrams also.

As a circumstellar shell develops, the M stars at first show redder [7] --
[12] colours, in relation to $M_{[12]}$ or $K_{S,0}$ -- [7], than the carbon
stars, which causes them to follow separate paths in the corresponding
diagrams. However, as a thick oxygen-rich shell develops, the [7] -- [12]
colour ultimately decreases somewhat. The optical thickness at 7$\mu$m is
probably still developing with increasing dust quantity even as it becomes
optically thick at 12$\mu$m. The 9.7$\mu$m silicate dust feature also
absorbs some of the flux in the [12] band under these conditions.

\section{Mass-loss}

The photospheric radiation at the longest wavelengths (12 or 15$\mu$m) can
be estimated by assuming that the $K$ fluxes are little affected by
circumstellar shells and extrapolating them as Rayleigh-Jeans
tails. The existence of the tight $K$--log$P$ relation for both C- and
O-rich stars suggests that $K$ is likely to be almost purely photospheric.

\begin{figure}
\epsfxsize=8cm
\epsffile[28 -9 453 786]{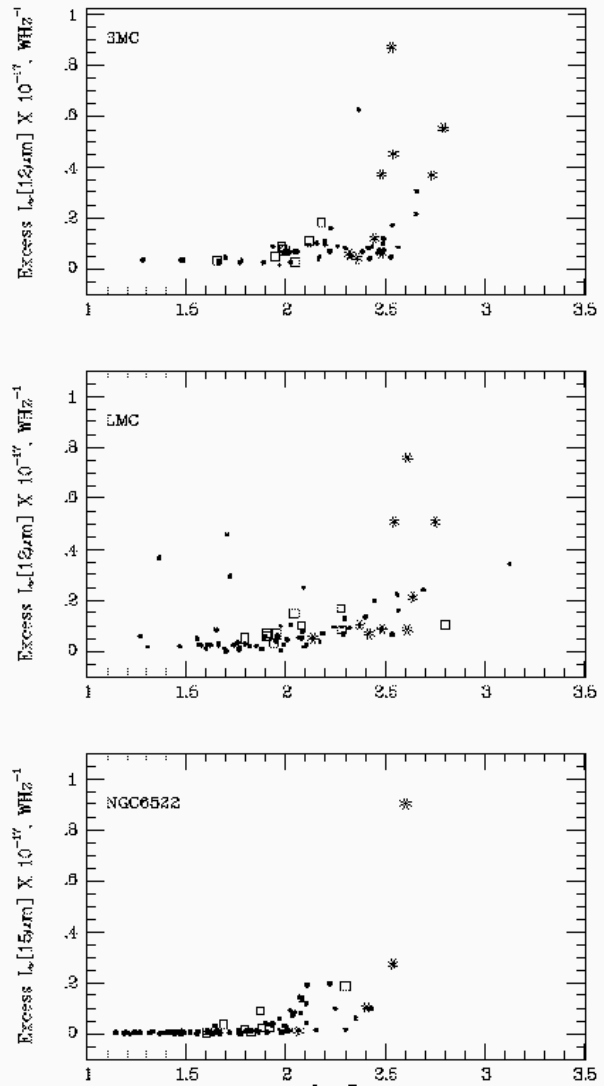}

\caption{Absolute monochromatic luminosities of the excess radiation
presumed to arise from dust formed during mass-loss, at the wavelengths
indicated, above the expected photospheric contribution, assuming that the
$K$ flux is all photospheric. Note the general similarity of mass-loss
levels (assumed to be affected only moderately by the difference between the
12 and 15$\mu$m filter wavelengths). The outlying semi-regular variable
points in the LMC fields are attributable to higher than usual errors in the
photometry.}

\label{XSKvsP.ps}
\end{figure}

The absolute monochromatic luminosity presumed to arise from dust emission
in circumstellar shells is plotted as a function of period in fig
\ref{XSKvsP.ps}. The mass-loss levels appear to be similar in spite of the
metallicity differences. All fields seem to show some stars with significant
mass-loss, beyond $P$ $\sim$ 100d. The double-period stars also seem
to stand out slightly as more likely to show an excess than variables of
similar short period. This is true even for presumably M stars in the
NGC 6522 field of the Galaxy. These diagrams will be affected strongly by
zero-point errors in the long-wavelength photometry.

\section{Conclusions}

The three fields we have examined show various similarities and differences.
Examining their near infrared colours and magnitudes we see that the level
of the RGB tip, $M_{K_{0}}^{\rm Tip}$ and the $J-K_S$ colour at a defined
$M_{K_S}$ (-- 5.5) are sensititive to metallicity as expected from galactic
globular cluster studies. Near the tip of the AGB the relative numbers of
luminous stars, often showing red $J-K_S$ colours, increases towards the
low-metallicity fields.

The variable star content of the fields also shows marked differences.
Although the same variability classes are present in all three fields, the
proportion of stars that are variable, as well as their amplitudes,
decreases towards lower metallicity. The period-luminosity loci in the $K$,
log$P$ diagram are also populated in different ways in the three fields.

The mid-IR observations are affected by the sensitivity limit of ISO and the
tendency towards larger relative numbers of luminous stars with
low-metallicity. The colour-colour and colour-magnitude diagrams can be used
to differentiate between oxygen- and carbon-rich stars with developed
circumstellar shells. The rates of mass-loss, as judged by the excess fluxes
at long wavelength, appear to be similar for the stars in each field.

\section{Acknowledgments}

ISG thanks the Institut d'Astrophysique de Paris for their hospitality
during part of this work. This visit was supported by the CNRS-NRF
agreement. He also thanks the Visitor Programme at the European Southern
Observatory for support during a visit of one month.

MS is supported by the APART programme of the Austrian Academy of Science
and visited the South African Astronomical Observatory for two weeks under
the CNRS-NRF agreement.

Cecile Loup (Institut d'Astrophysique de Paris), and Martin Groenewegen
(Institute of astronomy, Katholieke Universiteit, Leuven) are thanked for
discussion and information.

This paper utilizes public domain data originally obtained by the MACHO
Project, whose work was performed under the joint auspices of the U.S.
Department of Energy, National Nuclear Security Administration by the
University of California, Lawrence Livermore National Laboratory under
contract No. W-7405-Eng-48, the National Science Foundation through the
Center for Particle Astrophysics of the University of California under
cooperative agreement AST-8809616, and the Mount Stromlo and Siding Spring
Observatory, part of the Australian National University.

This publication makes use of data products from the Two-Micron All-Sky
Survey, which is a joint project of the University of Massachusetts and the
Infrared Processing and Analysis Center, funded by the National Aeronautics
and Space ASdministration and the national Science Foundation (USA).

{}

\end{document}